\documentclass[preprint,amsmath,amssymb]{revtex4}
\pdfoutput=1
\usepackage{graphicx}
\usepackage{dcolumn}
\usepackage{bm}
\usepackage{epstopdf}
\begin{document}

\title{Abatement of mixing in shear-free elongationally unstable viscoelastic microflows}
\author{R.M. Bryce}
\email{rbryce@phys.ualberta.ca}
\author{M.R. Freeman}%
\email{mark.freeman@ualberta.ca}
\affiliation{%
Department of Physics, University of Alberta, Edmonton T6G 2G7, Canada.\\
National Institute for Nanotechnology, 11421 Saskatchewan Drive, Edmonton T6G 2M9, Canada.
}%

\date{\today}

\begin{abstract}
The addition of minute amounts of chemically inert polyacrylamide polymer to liquids results in large instabilities under steady electro-osmotic pumping through 2:1 constrictions, demonstrating that laminar flow conditions can be broken in electro-osmotic flow of viscoelastic material. By excluding shear and imposing symmetry we create a platform where \emph{only} elongational viscoelastic instabilities, and diffusion, affect mixing. In contrast to earlier studies with significant shear that found up to orders of magnitude increase in mixing we find that inclusion of polymers excites large viscoelastic instabilities yet mixing is \emph{reduced} relative to polymer-free liquids. The absolute decrease in mixing we find is consistent with the understanding that adding polymer increases viscosity while viscoelastic flows progress towards elastic turbulence, a type of mild (Batchelor) turbulence, and indicates that electro-osmotic pumped devices are an ideal platform for studying viscoelastic instabilities without supplementary factors. 
\end{abstract}


\maketitle

\section{Introduction}

This paper explores how mixing is affected by  adding high molecular weight polymers to electro-osmotically pumped microflows. Adding low concentrations of polymer to fluids moderately increases viscosity and dramatically increases elasticity, allowing microflows to  undergo viscoelastic instability. Based on the relatively small increase in viscosity, the striking qualitative change in stability, and early experiments that demonstrate up to orders of magnitude improvement in mixing there is an expectation of dramatic increase in mixing by incorporation of polymers. By excluding shear, present in all mechanically pumped cavity flows, we find that contrary to such expectations absolute mixing is reduced by adding polymer to induce viscoelastic instability. Our findings demonstrate that viscoelastic instabilities do not, in themselves, promote mixing and suggest that electro-osmotic flows are an ideal platform for isolating and studying behavior of viscoelastic instabilities and elastic turbulence.

Viscoelastic flows are of practical interest for lab on a chip applications as many biological materials of interest for biomedical testing, such as saliva \cite{Hfilter} and blood \cite{blood}, are viscoelastic in nature as are many other suspensions \cite{Larson}. Several devices sort particles using laminar flow interactions \cite{micro_review}, suggesting that viscoelastic induced instabilities can have potential undesirable effects as more concentrated and complex fluids are investigated; it has already been noted that viscoelasticity in saliva testing negatively effects performance \cite{Hfilter}. In addition to the detrimental effects of instabilities there is suggestion that enhanced mixing may occur by intentionally spiking samples with polymers \cite{eturb_mix}, thereby causing instability and subsequently increased mixing. As chemical reaction rate and completion depends on good mixing and as many clinical, environmental, pharmaceutical, and other diagnostic applications require rapid and complete reactions, in particular in the presence of small concentrations of sample, efficient mixing is an important aspect to device performance. 

In microfluidic devices the scales involved (typically O(100) $\mu$m) are such that diffusive processes are slow, yet inertial fluid instabilities are suppressed due to the small scales leading to small Reynolds numbers. This intermediate ``slow scale" regime makes mixing a key issue in microfluidics \cite{umixreview} and various methods have been investigated in order to enhance mixing, including using bubbles \cite{bubblemix}, electrohydrodynamic instability \cite{EKI}, flow focusing \cite{squeezemix}, and bas-relief patterning \cite{herringbone}. These approaches generally rely on creating inhomogeneities to the working fluid or the microchannels. Viscoelastic fluids can exhibit flow instabilities \cite{Bird} when polymer coils stretch and bend along stream lines which creates hoop-stress, allowing polymers to cross steam lines and thereby disrupting laminar flow \cite{hoop}. 

These viscoelastic instabilities progress to elastic turbulence \cite{eturb}, a recently established turbulence that arises from elastic effects that can be excited at arbitrarily low Reynolds numbers. Use of viscoelastic fluids to promote mixing has been demonstrated in both microscaled Dean flows \cite{microflow}, where chaotic flow was established and where earlier study of a macroscaled version of the serpentine flow geometry demonstrated true elastic turbulence \cite{eturb_mix}. The establishment of viscoelastic turbulence \cite{eturb} and efficient mixing \cite{eturb_mix} along with favorable scaling properties of viscoelastic instabilities, which become easier to excite with reducing dimensions, suggests that the use of viscoelastic fluids may be a promising route to promote mixing in microfluidic devices. Furthermore, it is well know that solution viscosity increases with polymer concentration and at the overlap concentration, where neighboring coils impinge on each other, viscosity is only double that of the polymer-free solution \cite{microflow}. In contrast to this relatively mild quantitative change in viscosity (and thus decrease in diffusion) elasticity effects have dramatic qualitative impact with polymer concentrations as low as $\approx$ 10 ppm leading to both large drag reduction in high Reynolds number hydrodynamic turbulent flows \cite{polydrag} and sizable instability and elastic turbulence in low Reynolds number flows \cite{eturb_mix}.

Several studies (see, for example, \cite{microflow, xchannel, zigzag}) have found that viscoelastic instabilities can be excited in microdevices, however additional deformations of the flows - most notably shear - exists in prior work and highly viscous solvents are used, which is atypical for many lab on a chip applications. Here we use electro-osmotic flow (EOF) in order to exclude shear and solely induce elongational viscoelastic instabilities by driving low viscosity fluids through sudden 2:1 constrictions in linear microchannels; we compare mixing in unsteady viscoelastic flows relative to mixing in polymer-free solutions in order to measure mixing effects.

Mechanically driven cavity flows always have a significant shear component to the flow due to no-slip conditions. In contrast EOF has effective slip boundary conditions and shear is confined to the Debye layer close to the channel boundary \cite{micro_review}; flows are shear-free external to the Debye layer as the electric fields that the flow follows are purely elongational \cite{shear_free_E}. Further, for large polymers, entropic repulsion from the walls reduces interactions of the Debye shear regime with polymers \cite{dragEOF}. Our pumping scheme therefore excludes the shear component that is ubiquitous to mechanically pumped flows. In addition to deformations caused by shearing there are possible Lagrangian chaos effects \cite{Lchaos}, where flow fields result in chaotic trajectories which can lead to rapid mixing. We stack identical constriction units equally spaced down a microchannel, imposing symmetry as to reduce possible Lagrangian chaos effects \cite{pattern_chaos}. 

\begin{figure}
\includegraphics[scale=0.35]{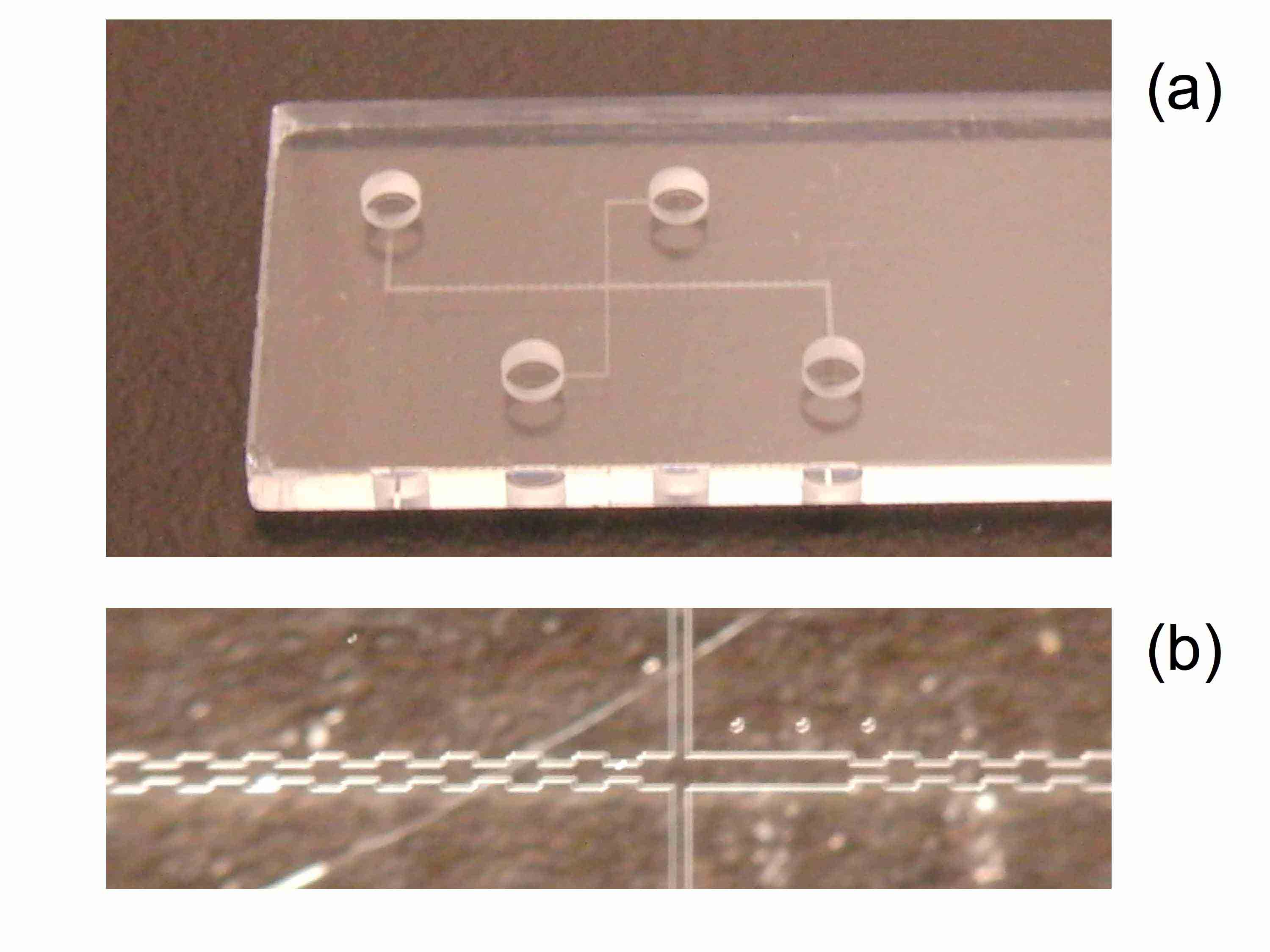}
\caption{\label{fig:chip}  (a) Photograph of the microfluidic device, and (b) detail of the microchannels with mixing units. We electro-osmotically drive fluids from the top and bottom wells to the left side well. Fluid in the top well contains fluorescent dye, and mixing is optically observed by exciting the dye with an external 532 nm laser and collecting video data. Each 100:200 $\mu$m constriction/expansion unit creates extensional flow which induces instabilities in viscoelastic fluids with high enough Deborah number. Due to the slip conditions of electro-osmotic flow and symmetry non-viscoelastic instability mixing mechanisms are minimized. We find flow rates are not affected by the presence of polymer in the dilute regime and drive all flows with a voltage of 200 V, resulting in equal flow rates and comparable flows.}
\end{figure}

\section{Experimental Setup}  

See Ref. \cite{PAA_EOF} were we demonstrate viscoelastic instabilitites under steady electro-osmotic pumping and discuss our device in more detail. Briefly, the experimental details are as follows: a microfluidic toolkit ($\mu$TK) from MicraLyne Inc. \footnote{See Ref.~\cite{uTKCrabtree} for a brief overview and discussion of the $\mu$TK apparatus.} is used to impose EOF flow in glass microchannels etched 20 $\mu$m deep. TAMRA dye, a fast diffusing dye with D $\approx$ 270 $\mu$m$^2$/s \cite{Boukari07}, fluorescently labels one, otherwise identical, stream of fluid allowing visualization of flows. A dyed and an un-dyed fluid stream are brought together in a cross-slot junction and flow downstream through 200:100 $\mu$m constrictions (see Fig.~\ref{fig:chip} for images of the device and microchannels). Fluid flow details are captured using a a Casio Exilim EX-F1 camera in video mode (30 fps), by attaching the camera to the $\mu$TK via a custom machined attachment and a Scoptronix Maxview Plus optical adaptor (camera-microscope attachment). 

Our camera is not scientific grade and comparison of CCD intensity with true (incident) laser power reveals nonlinear characteristics; cubing the CCD values results in a linear response profile allowing the optical intensity as measured to be scaled and related to the dye concentration. From video data space-time (xt) diagrams are created by taking a vertical cut from each video frame across the channel, horizontally centered in the constriction or expansion. Due to chemical etching side walls exhibit curvature, varying the channel depth near the edges, and we remove 20 $\mu$m nearest the channel edges before analysis. Runs are nominally 100 s in duration, but we exclude initial transient flow to prevent start up effects from influencing results.

Polymer solutions were made by adding high molecular weight ($18\times10^6$ Da) polyacrylamide from Polysciences Inc. to a 20:80 vol.\% methanol:water mixture; methanol is used to prevent aging during storage \cite{PEOdecay}. The overlap concentration $c_{*}$ is $\approx$ 300 ppm, and we work in the dilute regime ($c<c_{*}$). 

Flow velocities in EOF are given by $v=\mu_{\textrm{EOF}}E$, where $\mu_{\textrm{EOF}}$ is the electro-osmotic mobility and $E$ the applied electric field. We find that below the overlap concentration the electro-osmotic mobility is not significantly modified from the pure solvent case, with $\mu_{\textrm{EOF}} \approx (5.6\pm0.5) \times10^{-4}~\textrm{cm}^2/\textrm{Vs}$ here, a result that is predicted by the entropic wall exclusion effect and studied in Ref. \cite{dragEOF}. Our flow velocities are on the order of mm/s, and Re $<<$ 1, as is typical of standard EOF microflows.

\section{Results and Discussion}

Addition of polymer allows elastic effects to induce instabilities for sufficiently high Deborah number De above the instability threshold ($>$ 1/2 for elongational flows \cite{coil_transition_elongational}), the dimensionless number that characterizes viscoleastic flows. The Deborah number is given by

\begin{equation}
         De=\frac{\lambda v}{L_{\textrm{Char}}} 
              \approx \frac{\lambda_{Zimm} \mu_{\textrm{EOF}} E}{L_{\textrm{Char}}}.
 \label{eq:De}
\end{equation}
          
We estimate the relaxation time $\lambda$ of our liquid using the Zimm estimate $\lambda_{Zimm} \approx$ 0.03 s and $L_{\textrm{Char}} \approx 12~\mu$m for our channels; see Ref. \cite{PAA_EOF}. For our setup and polymer De $>1$/2 for driving voltages $\gtrsim$ 20 V, and here we operate well above the threshold at 200 V to ensure highly unstable flows. As we find $\mu_{\textrm{EOF}}$ is not modified by the presence of polymer in the dilute concentration regime all flows are comparable.

The strategy to promote instabilities discussed here is simple. The addition of polymer is facilitated if pre-solvated polymer mixtures are added to working fluids. The use of high viscosity solvents or concentrated polymer is not required: low viscosity, lightly polymer doped liquids undergo instability under typical electro-osmotic flow conditions.

The mixing efficiency is found by measuring the homogeneity of a fluid domain \cite{Danck52}, and here is calculated using  the first moment mixing index

\begin{equation}
\label{eq:mix_moment}
	   \textrm{M}1= \frac{\langle |c-\overline{c}| \rangle}{\overline{c}}
\end{equation}	

where c is the dye concentration and $\overline{c}$ the average concentration; for unmixed streams M1 = 1 and mixing is perfect for M1 = 0 (homogenous). 

\begin{figure}
\includegraphics[scale=0.32]{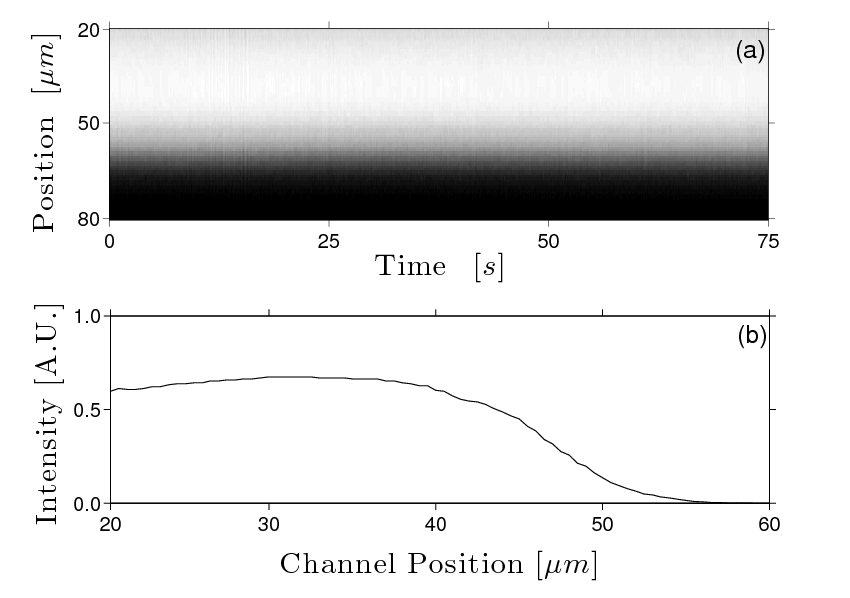}
\caption{\label{fig:laminar_N1} Laminar flow in the first constriction. With no polymer added creeping laminar flow with steady diffusion is observed (a). A space time diagram (a) is constructed by taking a cross channel intensity profile across the middle of the constriction and is 60 $\mu$m by 75 s. Note that 20 $\mu$m is trimmed from the edges (where wet etch rounding may lead to aberrations), and that transitory start-up flow is removed, from all xt-diagrams here. The cross channel intensity profile 0.6 s from the start of the xt diagram is shown in (b). An erf-like profile is observed in (b), as expected in coflowing two-stream mixing.}
\end{figure}

In ``normal" microfluidics smooth laminar flows are observed, as seen in Fig.~\ref{fig:laminar_N1} for our polymer-free solutions. As the fluid flows downstream diffusion across the channel will eventually result in homogenously mixed fluid (not show). We observe that mixing values for polymer-free flows scale approximately exponentially as a function of distance downstream; while perfect exponential scaling is not expected calculating M1 using the theoretical erf-profile \cite{Crank75} for laminar mixing in uniform channels results in an approximate exponential drop off \footnote{The linear norm of residuals is $<$ 0.01 over the range of mixing units tested, indicating exponential scaling is a close approximation of mixing behavior here.}.

\begin{figure}
\includegraphics[scale=0.32]{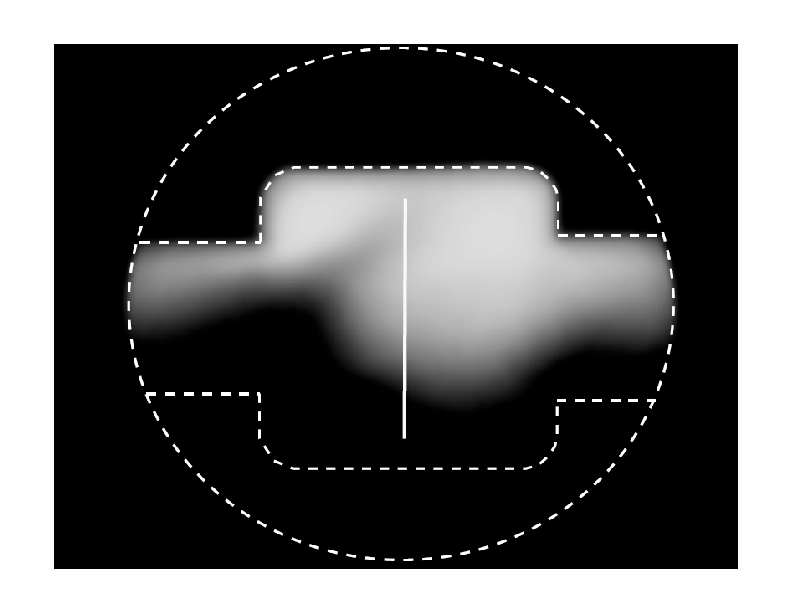}
\caption{\label{fig:photo_ex_con}Photomicrograph of unstable flow in the second expansion for a concentration of 64 ppm high molecular weight ($18\times10^6$ Da) polyacrylamide polymer added. It is apparent that large instabilities are excited that display a dominant scale set by the microchannel. Dashed outlines indicate field of view and microchannel outline; the 160 $\mu$m long solid line indicates where the space-time diagram slice was taken for the expansion regions. The instabilities in the constriction regions can also be seen here at the sides of the image; regions are centered for data collecting.}
\end{figure}

\begin{figure}
\includegraphics[scale=0.32]{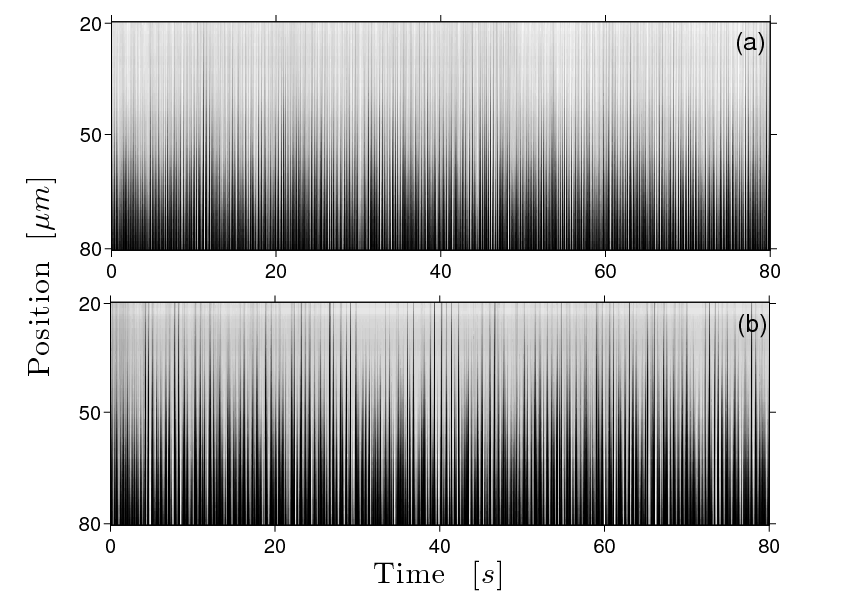}
\caption{\label{fig:mixing_N2}Space-time diagrams of unstable flow in the second constriction. Adding 32 ppm (a) and 64 ppm (b) polymer results in instabilities as fluid flows through the constriction/expansion units. Visual comparison indicates that instabilities more rapidly fluctuate across the channel for lower concentration, and that instabilities move more material across the channel with increased polymer concentration. The diagrams are 60 $\mu$m by 80 s, and have been scaled here to increase the contrast.}
\end{figure}

\begin{figure}
\includegraphics[scale=0.32]{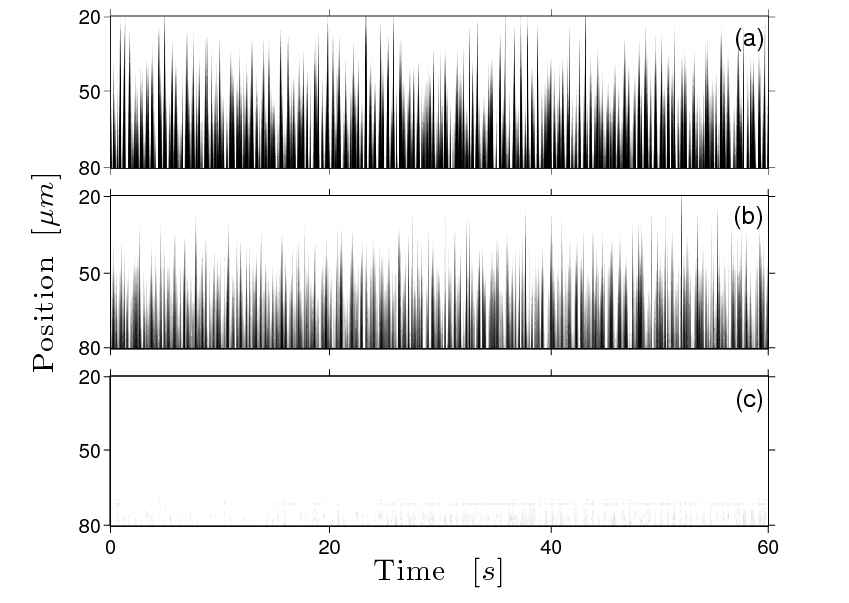}
\caption{\label{fig:mixing_fn_N} Mixing evolution. From the top to the bottom 64 ppm sample passes from the 2$^{nd}$ (a) to the 4$^{th}$ (b) to the 8$^{th}$ (c) constriction. The effects of diffusion are apparent, with a reduction of contrast between the 2$^{nd}$ (a) and the 4$^{th}$ constriction (b), and finally to a visually homogeneous mixture by the 8$^{th}$ constriction (c). In addition it can be seen that strong cross channel fluctuations exist, but the pattern is highly similar and a comb-like pattern persists between the 2$^{nd}$ (a) and the 4$^{th}$ (b) constriction, with addition of ``teeth" but without significant pattern changing deformation. This persistence is characteristic of chaotic flows undergoing symmetric oscillatory forcing. The diagrams are 60 $\mu$m by 60 s and have not been scaled, allowing comparison.}
\end{figure}

Addition of polymer results in large instabilities, as can be seen in Fig.~\ref{fig:photo_ex_con} and Fig.~\ref{fig:mixing_N2}. The fluctuations move material transverse to the flow direction, spanning almost the entire channel. There is no dominant frequency in the flow, indicating chaotic flow \cite{microflow}, and the overall pattern displayed are striations with ``comb-like" teeth randomly spaced. Increasing the concentration from 32 to 64 ppm results in larger fluctuations and tooth spacing. As the solutions move downstream through additional elongation regions additional striations are added, as seen in Fig.~\ref{fig:mixing_fn_N}, however the general pattern persists with diffusion appearing to be the dominant mixing mechanism.

\begin{figure}
\includegraphics[scale=0.32]{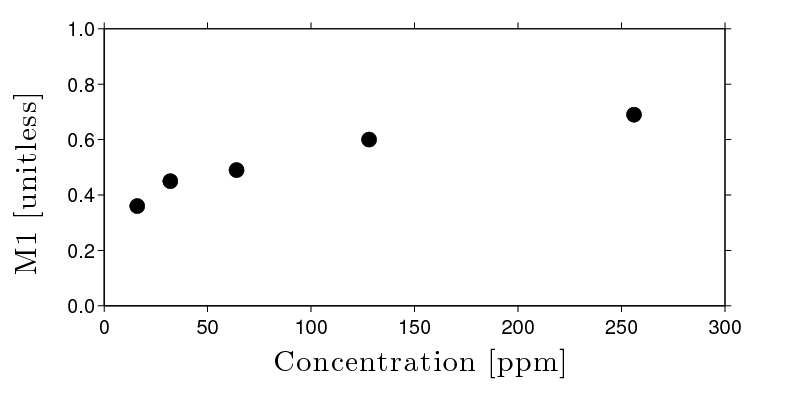}
\caption{\label{fig:mix_c}  Mixing as a function of concentration in the 2$^{nd}$ expansion unit, as measured by the first moment mixing index M1 (M1 $\rightarrow$ 0 for perfect mixing and M1 $\rightarrow$ 1 for no mixing). Using space-time diagrams mixing is measured for c = 16, 32, 64, 128, and 256 ppm (the overlap concentration is $c_{*} \approx$ 300); error in M1 is estimated as $\lesssim$ 5\%. The mixing index slowly raises with concentration in the dilute regime; this indicates that while instabilities are promoted by the presence of polymers \emph{absolute} mixing is reduced (diffusion is hindered by the presence of polymers). The lack of enhanced diffusion demonstrates that small scale instabilities are not excited to any significant extent.}
\end{figure}

Absolute mixing is reduced by reduced diffusion, and increased by instability induced deformation of the liquid. There is therefore a tradeoff involved in adding polymer to the solution which will allow viscoelastic instabilities yet reduces diffusion. By measuring mixing (M1) in the second expansion we find that mixing is reduced as concentration is increased (see Fig.~\ref{fig:mix_c}), demonstrating that the effect of instabilities in promoting mixing is not dominant for mixing our dye. As our dye is a small molecule ($\approx 1.6$ nm hydrodynamic diameter \cite{Boukari07}) our mixing experiments is a stringent - but realistic - test. Further, as we trim the edges of the xt-diagrams to reduce image processing artifacts we possibly artificially increase the first moment mixing index M1 calculated for unstable flows, as material with high/low dye concentration can be injected from these regions, increasing the variance seen. 

Note that in contrast to high Reynolds number turbulent flows an enhanced diffusion rate, indicative of broad-band excitation of instability scales, is not observed \cite{taylor_diffusion} supporting the limited bandwidth visual evidence (see Fig.~\ref{fig:photo_ex_con}) that large scale instabilities dominate the instability spectrum. The lack of enhanced diffusion confirms that small scale instabilities below the camera resolution are not excited to a significant extent.

\begin{figure}
\includegraphics[scale=0.32]{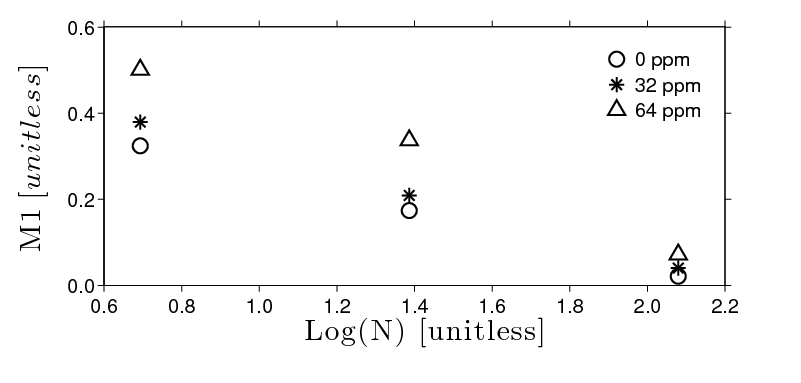}
\caption{\label{fig:mixing_c}  Mixing as a function of constriction unit, as measured by the first moment mixing index M1. Using space-time diagrams mixing was measured downstream of the initial junction at N=2, 4, 8. The mixing index scales exponentially, and the mixing lengths mildly reduce with concentration (4.5 for 0 ppm, 4.2 for 32 ppm, and 3.2 for 64 ppm). The  vertical shift for polymer doped fluids relative to undoped fluid is due to reduced diffusion limiting the mixing, as increased polymer concentration reduces the diffusion constant by increasing viscosity.}
\end{figure}

Finding the first moment mixing index, M1, for concentrations of 0 ppm, 32 ppm, and 64 ppm as the fluid moves downstream reveals that  absolute mixing is reduced in viscoelastic solutions relative to the polymer-free sample; Fig.~\ref{fig:mixing_c}. We also observe an increase in the rate of change in mixing downstream for viscoelastic solutions. By looking at the flow at different constriction units (N=2, 4, 8) the mixing length is calculated and we find a mild decrease with raising polymer concentration; the mixing length reduces from 4.5 for 0 ppm to 4.2 for 32 ppm and finally 3.2 for 64 ppm. For diffusion from an ideal sharp 2D interface \cite{Crank75} changing viscosity simply shifts calculated M1 by a constant for our parameters, indicating that the change in mixing length can be attributed to the action of instabilities. 

For an uniform microchannel we calculate \cite{Crank75} M1 values roughly twice as large as we observe for polymer-free solutions in our channels corrugated with constriction/expansions. This enhanced mixing has been previously observed for electro-osmotic flows through constrictions/expansion units where $\approx$ 2$\times$ enhancement was found \cite{EOF_mixers}. This enhancement is an open problem currently being investigated \cite{transport_varies_with_channel}, but is thought to be due to the modulation of width enhancing diffusion \cite{transport_varies_with_channel} due to convective cross flow movement (e.g. periodic movement perpendicular to the main flow direction) \footnote{Naively considering the sheaves of flow lines suggests that the total distance an average particle travels as it moves downstream in a corrugated channel will be larger than the minimum distance (e.g. channel length); fluid off center will effectively ``see" a longer channel and thus experience more diffusion, and one should expect enhanced overall diffusion in a given length relative to the centerline flow.}. 

We note that Lam \emph{et al.} Ref. \cite{whip} create instabilities by driving viscoelastic fluids through abrupt 8:1 constrictions using a syringe pump, and therefore have a mechanically driven device similar to our device. They observe enhancement of mixing in their liquids over deionized water at equal flow rates, demonstrating that \emph{absolute} mixing increase can be obtained over low viscosity Newtonian liquids. However, they use dis-similiar fluids of different viscosity, shear-thinning, and elastic properties making interpreting their results difficult as flow focusing, shear-thinning \cite{Kaye}, and viscous thread folding \cite{threads} or other coflowing effects are physical mechanisms that all may play important roles in addition to viscoelastic instabilities in promoting the mixing they observe.

We have shown that viscoelastic instabilities alone will not lead to rapid mixing, in contrast to earlier studies \cite{eturb_mix, microflow, zigzag} that found up to orders of magnitude improvement in mixing for mechanically driven channel flow. However, significant shear stretching is present in most prior work due to no-slip conditions. In a study of electrokinetic driven 2D sheet flow \cite{poly_mix_2D}, which also excludes no-slip induced shear, viscoelasticity was found to mildly suppress mixing for flows below the instability threshold, a finding that is consistent with other studies that find either mild increase \cite{poly_chaos_increase} or decrease \cite{poly_chaos_decrease} in this regime where polymers are only weakly distorted. Additionally, in Ref. \cite{poly_mix_2D} they show that above the instability threshold stretching fields are reduced in viscoelastic fluids. By comparing a Newtonian and a highly shear-thinning/viscoelastic fluid they find enhanced stretching and mixing for the shear-thinning/viscoelastic fluid, indicating that mixing trends are correlated to stretching field magnitude. Our finding of reduced mixing, combined with the reduced stretching fields observed in Ref. \cite{poly_mix_2D}, shows the correlation between stretching fields and mixing holds for viscoelastic fluids above the instability threshold. Our results which demonstrate reduced mixing in unstable viscoelastic flows are consistent with the understanding that viscoelastic fluids driven well above the viscoelastic instability threshold progress towards elastic turbulence which has been shown to be a realization of Batchelor turbulence \cite{eturb_archival, vKeturb}.

In Batchelor turbulence it has been established that the velocity spectrum decays rapidly (the wavenumber $k$ scales as $k^{-x}$, where $x$ has experimentally been found to be in the range 3.3 to 3.6 \cite{vKeturb}) which leads to weak coupling between scales, in sharp contrast to high Reynolds number turbulence, and the resulting flows are random in time yet spatially smooth and strongly correlated on the integral scale \cite{vKeturb}. This results in instabilities dominated by large scale fluctuations, which are set by the system size \cite{vKeturb}, and therefore there will remain a mismatch between the smallest scale one wants homogeneity on and the instability which is intended to drive mixing. Note that the smooth flow characteristics of Batchelor turbulence are consistent with our observations (see, for example, Fig.~\ref{fig:photo_ex_con}) and, without some other mechanism (such as shear) to further modify flow, are at odds with enhanced mixing.

To create efficient mixing a bridge is needed between the relatively large scale set by viscoelastic instabilities and the smallest scales over which homogeneity is required; diffusion, Lagrangian chaos, or deformations such as shear can bridge these scales. In the ``knead and fold" metaphor of mixing \cite{LTM_book} viscoelastic instabilities can provide the fold operation that creates large deformations of material to create striations of dis-similiar material, however additional deformations are needed to thin striations allowing diffusion to finally homogenize the fluids.

\section{Conclusion} 

In summary, the addition of high molecular weight polyacrylamide to liquids leads to striking viscoelastic instabilities in extensional electro-osmotic flow thereby upsetting the laminar flow typical of microchannels. The breaking of laminar flow for dilute low viscosity viscoelastic fluids has implications for device design as laminar flow is often assumed in microflows; as lab on a chip applications move towards more complex fluids this assumption must be checked. Our results show that for flows of lightly polymer-doped low viscosity solutions through unoptimized geometries mixing is \emph{reduced} relative to polymer-free solution, despite the excitation of dramatic large scale viscoelastic instabilities. This reduction is in contrast to earlier studies that found up to orders of magnitude increase in mixing, demonstrating that the role of shear or other deformations of the flow is crucial in promoting mixing. Electro-osmotic flows can be used to isolate viscoelastic instabilities, and combined pressure/EOF driven flows \cite{EOF_P_poly} allow controlled application of shear enabling detailed investigation of flow evolution and mixing. In addition to fundamental studies possibility for application exists for the observed highly unstable, yet diffusion limited, flows in applications which depend on large interfacial area; for example, in membraneless fuel cells \cite{fuel_cell} or for use in combination with the liquid junction potential \cite{LJP}.

\begin{acknowledgments}
We are grateful for support from the National Institute for Nanotechnology, the Natural Science and Engineering Council of Canada, the informatics Circle of Research Excellence, and the Canada Research Chairs program. We thank CMC for fabrication of the devices. We would like to thank Zhe Wang for help in the laboratory and with data analysis, and Dave Fortin for creating the illustrated contents image.
\end{acknowledgments}

\end{document}